\documentclass[letter,12pt]{article}

\usepackage{amsmath}
\usepackage{amssymb}
\usepackage{hyperref}

\def\RCFT{\mathrm{RCFT}}
\def\SU{\mathrm{SL}}

\def\SU{\mathrm{SU}}

\def\U{\mathrm{U}}
\def\cN{\mathcal{N}}
\def\cA{\mathcal{A}}
\def\cB{\mathcal{B}}
\def\bC{\mathbb{C}}
\def\bR{\mathbb{R}}
\def\bZ{\mathbb{Z}}
\let\hat\widehat

\begin{document}

\begin{titlepage}

\bigskip

\begin{flushright}
\normalsize
PUPT-2382

\medskip

June, 2011
\end{flushright}

\bigskip

\begin{center}

{\LARGE 
Para-Liouville/Toda central charges from M5-branes
}
\end{center}

\vfil
\vfil
\medskip

\begin{center}
\def\thefootnote{\fnsymbol{footnote}}

Tatsuma Nishioka$^\heartsuit$ and  Yuji Tachikawa$^\clubsuit$\footnote[1]{on leave from IPMU, the University of Tokyo}
\end{center}

\begin{flushleft}\small

$^\heartsuit$ Department of Physics, Princeton University,  Princeton, NJ 08544, USA

$^\clubsuit$ School of Natural Sciences, Institute for Advanced Study,  Princeton, NJ 08540, USA

\end{flushleft}
\vfil
\bigskip

\begin{center}
{\bfseries abstract}
\end{center}

\bigskip

We propose that $N$ M5-branes,  put on $\bR^4/\bZ_m$ with deformation parameters $\epsilon_{1,2}$, realize two-dimensional theory with $\hat\SU(m)_N$ symmetry and $m$-th para-$W_N$ symmetry. 
This includes the standard $W_N$ symmetry for $m=1$ and super-Viraroro symmetry for $m=N=2$.
 We provide a small check of this proposal by calculating the central charge of the 2d theory from the anomaly polynomial of the 6d theory. 

\vfill

\end{titlepage}

\paragraph{Introduction:}
$N$ M5-branes, put on $\bR^4$ with Nekrasov's deformation parameters $\epsilon_{1,2}$, are now believed to realize two-dimensional theory with $W_N$ symmetry \cite{Alday:2009aq,Wyllard:2009hg}.
One check of this statement was given in \cite{Alday:2009qq} following  the observation made in \cite{Bonelli:2009zp}.
Namely, the equivariant integral on $\bR^4$ of the anomaly polynomial of $N$ M5-branes determined in \cite{Harvey:1998bx} gives the  anomaly polynomial of the 2d theory, from which the central charge of the Toda theory with $W_N$ symmetry can be reproduced. The same analysis can be performed for 6d $\cN=(2,0)$ theory of type $G=A,D,E$ whose anomaly polynomial is also known \cite{Intriligator:2000eq,Yi:2001bz}; and it correctly reproduces the central charge of the Toda theory of type $G$.

In a recent paper \cite{Belavin:2011pp}, it was proposed that two M5-branes on $\bR^4/\bZ_2$ with deformations $\epsilon_{1,2}$ give rise to a system with the $\hat\SU(2)_2$ symmetry and the super-Virasoro symmetry. 
As a generalization, we propose that  $N$ M5-branes on $\bR^4/\bZ_m$ realize a 2d system with a free boson, $\hat\SU(m)_N$, and the $m$-th para-$W_N$ symmetry.
Here $\bZ_m$ acts as $(z,w)\mapsto (e^{2\pi/m}z,e^{-2\pi/m}w)$ on $(z,w)\in\bC^2\simeq \bR^4$.
We give a small piece of supporting evidence by calculating the central charge from the 6d anomaly polynomial. We will also speculate what happens if the 6d $\cN=(2,0)$ theory of  type $G$ is used instead. In the following $G$ stands for one of $A_n$, $D_n$ or $E_n$; $r_G$, $h_G$ and $d_G$ are the rank, the (dual) Coxeter number and the dimension of $G$, respectively. They satisfy $d_G=r_G(h_G+1)$.

\paragraph{Para-W symmetry and para-Toda theory:}
One way to realize the $W(\hat G)$ symmetry \cite{Christe:1988vc,Bowcock:1988vs,Nam:1989ya,Ahn:1990gn,Bernard:1990ti,Nemeschansky:1991pr} is to consider the chiral algebra of the coset \begin{equation}
\hat G_k \times \hat G_{m} / \hat G_{k+m} \ ,
\end{equation} for $m=1$. The $m$-th para-$W(\hat G)$ symmetry is obtained by taking an arbitrary positive integer $m$ in this coset; in particular, it reduces to the super-Virasoro algebra when $m=N=2$. Generalization of NSR superstrings using these algebras were explored e.g.~in \cite{Argyres:1990aq,Irie:2009yb}.

The $m$-th para-$W(\hat G)$ algebra is the symmetry of the $m$-th para-Toda model of type $G$, 
which has the following action \cite{LeClair:1992xi} \begin{equation}
S=S\left(\frac{\hat G_m}{\hat \U(1)^{r_G}}\right) + \int d^2x \left[\partial_\mu \Phi \partial_\mu \Phi +  \sum_{i=1}^{r_G} \Psi_i \bar\Psi_i \exp\left(\frac{b}{\sqrt m}\, \alpha_i\cdot \Phi\right) \right] \ .
\end{equation} Here, $\hat G_m/\hat \U(1)^{r_G}$ describes the generalized parafermions $\Psi_i$ of type $G$ \cite{Gepner:1987sm}, $\alpha_i$ are simple roots of $G$, $\Phi$ are $r_G$ free bosons with background charge $(b+1/b)\rho/\sqrt{m}$ with the Weyl vector $\rho$. 
 The central charge is given by \begin{equation}
c= c\left(\frac{\hat G_m}{\hat \U(1)^{r_G}}\right)+r_G+ \frac{h_G d_G}{m}\left(b+\frac1b\right)^2=\frac{m d_G}{m+h_G} + \frac{h_G d_G}{m}\left(b+\frac1b\right)^2 \ .\label{para}
\end{equation}
Note that the parafermion $\Psi_i$ has dimension $1-1/m$, and the exponential of bosons has dimension $1/m$, so that the interaction terms are marginal.
For $m=1$ this is the usual affine Toda theory, and for $m=N=2$ this is the $\cN=1$ super-Liouville theory.

\paragraph{$N$ M5-branes on $\bR^4/\bZ_m$:}
The anomaly polynomial $I_8$ of 6d $\cN=(2,0)$ is given by the general form \begin{equation}
I_8 = \cA\, I_8(1) + \cB\, p_2(NW)/24 \ ,\label{a}
\end{equation} where $I_8(1)$ is the anomaly polynomial of a single M5-brane,  $NW$ is the normal bundle to the worldvolume $W$ of the 6d theory, and $\cA,\cB$ are integers determined by the type of the 6d theory. 
For $N$ M5-branes, $\cA=N$ and $\cB=N^3-N$.
When compactified on a four-manifold $X_4$ with a suitable twist, it was determined in \cite{Alday:2009qq} that the resulting 2d theory has the central charge \begin{equation}
c=\chi(X_4)\cA + (P_1(X_4)+2\chi(X_4))\cB \ .\label{b}
\end{equation}  Here $\chi(X_4)$ and $P_1(X_4)$ are the Euler number and three times the signature of $X_4$, respectively. We let $X_4$ be $\bR^4/\bZ_m$ with the deformations $\epsilon_{1,2}$. Then $\chi$ and $P_1$ are to be taken in the equivariant sense, and are given by\footnote{The calculation is done as follows. Let us parameterize $\bR^4\simeq \bC^2$ by $(z,w)$, on which two rotations act via $(z,w)\mapsto (e^{\epsilon_1}z,e^{\epsilon_2}w)$. On the blowup of $\bR^4/\bZ_m$, we have $m$ fixed points of $\U(1)^2$ actions, whose local coordinates are given by $(z_i,w_i)=(z^{m-i+1}w^{1-m},z^{i-m}w^{m})$ for $i=1,\ldots,m$. Let us define $\epsilon_{1,2}(i)$ by the $\U(1)^2$ action at the fixed points: $(z_i,w_i)\mapsto (e^{\epsilon_1(i)}z_i,e^{\epsilon_2(i)}w_i)$. Then the topological numbers are given by the fixed point formula: $\chi=\sum_i 1 $, and 
$P_1=\sum_i (\epsilon_1(i)^2+\epsilon_2(i)^2)/(\epsilon_1(i)\epsilon_2(i)). $ } 
\begin{equation}
\chi(X_4)=m \ , \qquad
P_1(X_4)=\frac1m\frac{(\epsilon_1+\epsilon_2)^2}{\epsilon_1\epsilon_2} -2m \ .
\end{equation}
Therefore, $N$ M5-branes on $\bR^4/\bZ_m$ give rise to a 2d system with the central charge \begin{equation}
c=Nm + \frac{N^3-N}{m}\left(b+\frac1b \right)^2 \ , \label{foo}
\end{equation} where we used the standard identification $\epsilon_1/\epsilon_2=b^2$.

According to the general lore, 
the Hilbert space of the 2d theory comes from the BPS states of the supersymmetric quantum mechanics on the moduli space of instantons on $X_4$. 
In our case, we consider $\U(N)$ instantons on $\bR^4/\bZ_m$,
for which it is known that there is an action of a free boson and of $\hat\SU(m)_N$;
this was found in \cite{Nakajima1,Nakajima2} and string theory interpretation was later given e.g.~in \cite{Dijkgraaf:2007sw}.
Then, the central charge \eqref{foo} needs to be subdivided to \begin{equation}
c=1+c\left(\hat\SU(m)_N\right) + \left[\frac{m(N^2-1)}{m+N} + \frac{N^3-N}{m}\left(b+\frac 1b\right)^2 \right] \ .\label{bar}
\end{equation} 
The third term is the central charge of the $m$-th para-Toda theory of type $\SU(N)$ we found in Eq.~\eqref{para}.

We interpret this 
calculation as a check to the proposal that $N$ M5-branes on $\bR^4/\bZ_m$ with deformations $\epsilon_{1,2}$ give rise to a 2d system with actions of a free boson, $\hat\SU(m)_N$  and $m$-th para-$W_N$ algebra with central charge \eqref{para}. This statement reduces to the now-standard relations in \cite{Alday:2009aq,Wyllard:2009hg} when $m=1$, and to the proposal in \cite{Belavin:2011pp} when $m=N=2$.

\paragraph{Speculation concerning 6d theory of general type on $\bR^4/\bZ_m$:}
First let us consider what happens if we start from 6d $\cN=(2,0)$ theory of type $A_{N-1}$, instead of $N$ M5-branes. One needs to decouple the center-of-mass mode, which changes $\cA,\cB$ in Eqs.~\eqref{a}, \eqref{b} to $\cA=N-1$ and $\cB=N^3-N$. The resulting 2d theory has the central charge of the form \begin{equation}
c=c\left(\frac{\hat\SU(m)_N}{\hat\U(1)^{m-1}}\right) + \left[ c\left(\frac{\hat\SU(N)_m}{\hat\U(1)^{N-1}}\right)+(N-1) + \frac{N^3-N}{m}\left(b+\frac 1b\right)^2 \right]\ .
\end{equation} We see two cosets realizing generalized parafermions, known also as $\RCFT[A_{m-1},A_{N-1}]$ and $\RCFT[A_{N-1},A_{m-1}]$ in the terminology of \cite{Cecotti:2010fi}, respectively. In general, $\RCFT[\Gamma,G]$ for $\Gamma,G=A,D,E$ is a rational CFT with central charge \begin{equation}
c\left(\RCFT[\Gamma,G]\right)=\frac{h_{\Gamma}r_\Gamma r_{G} }{h_\Gamma+h_G} \ .
\end{equation} 
Note that $\RCFT[G,A_{m-1}]$ is the generalized parafermion of type $G$, but that
$\RCFT[\Gamma,G]$  is not yet constructed for general pair of $G$, $\Gamma$. 
We have \begin{equation}
c\left(\RCFT[\Gamma,G]\right)+c\left(\RCFT[G,\Gamma]\right)=r_\Gamma r_{G} \ ,
\end{equation} and it is believed they comprise a `level-rank-dual' pair of RCFTs.

The 6d theory of type $G$ has $\cA=r_G$ and $\cB=d_G h_G$ in Eqs.~\eqref{a}, \eqref{b} \cite{Intriligator:2000eq,Yi:2001bz}. Using $d_G=r_G(h_G+1)$, we find that the 6d theory on $\bR^4/\bZ_m$ has  the central charge \begin{equation}
c=c\left(\RCFT[A_{m-1},G]\right)+\left[ c(\RCFT[G,A_{m-1}])+r_G+\frac{d_G h_G}{m}\left(b+\frac1b\right)^2 \right] \ ,
\end{equation} where the second term is the central charge of the $m$-th para-Toda theory of type $G$ we saw in Eq.~\eqref{para}.
This suggests that the resulting 2d theory has the symmetry $\RCFT[A_{m-1},G]$ and the $m$-th para-$W(\hat G)$ symmetry.  Note that $\RCFT[A_{m-1},G]$ is not yet constructed when $G\ne A$.

We can further generalize the system by considering instantons of gauge group $G=A,D,E$ on the ALE orbifold of type $\Gamma$. Nekrasov's deformation cannot be performed when $\Gamma\ne A$, because the ALE orbifold of type $D$ and $E$ does not have $\U(1)^2$ isometry. We can still expect that this construction might naturally give us  $\RCFT[\Gamma,G]+\RCFT[G,\Gamma]$. The symmetry under $G$ and $\Gamma$ can be understood once one realizes that the 6d theory of type $G$ is Type IIB string on the ALE orbifold of type $G$. Then, the 2d system is Type IIB string on the ALE orbifold of type $G$ times the ALE orbifold of type $\Gamma$, which is manifestly symmetric under the exchange of $G$ and $\Gamma$.

\paragraph{Acknowledgments:}
The authors are extremely grateful to H.~Irie for his advise on everything concerning the parafermionic Liouville/Toda theories. Without him the analysis given in this paper was surely not possible.
The authors also thank D.~Gaiotto, K.~Maruyoshi, G.~Moore and M.~Yamazaki for discussions.
TN is supported in part by NSF grants PHY-0844827 and PHY-0756966.
YT is supported in part by NSF grant PHY-0969448  and by the Marvin L. Goldberger membership through the Institute for Advanced Study.
YT is also supported in part by World Premier International Research Center Initiative (WPI Initiative),  MEXT, Japan through the Institute for the Physics and Mathematics of the Universe, the University of Tokyo.

\bibliographystyle{ytphys}
\small
\baselineskip=.8\baselineskip
\bibliography{fractional}{}

\end{document}